\documentstyle[aps]{revtex}
\def \be   {\begin{equation}}
\def \ee   {\end{equation}}
\def \l {\label}
\begin{document}
\input epsf
\title{A new approach to gauge fields}
\author{Manoelito M de Souza}
\address{Universidade Federal do Esp\'{\i}rito Santo - Departamento de
F\'{\i}sica\\29065.900 -Vit\'oria-ES-Brasil}
\date{\today}
\maketitle
\begin{abstract}
A discrete field formalism exposes the physical meaning and the origins of gauge fields and of their symmetries and singularities.
\end{abstract}
\begin{center}
\end{center}

The Maxwell's equations 
\be
\l{hM}
\varepsilon^{\mu\nu\rho\sigma}\partial_{\nu}F_{\rho\sigma}=0,
\ee
\be
\l{iM}
\partial_{\nu}F_{\mu\nu}=4\pi J^{\mu},
\ee
contain all the accumulated knowledge about the electromagnetic fields $F$, their sources $J$, and their interactions. The homogeneous set of equations has 
\be
\l{hs}
F_{\mu\nu}=\partial_{\mu}A_{\nu}-\partial_{\nu}A_{\mu},
\ee
as a generic solution that turns the second set of equations into
\be
\l{iMa}
\Box A^{\mu}-\partial^{\mu}\partial.A=-4\pi J^{\mu},
\ee
which is not solvable because the four equations cannot be disentangled. But only $F$, and not $A$, is directly  accessible to experimental detection, and $F$ is an antisymmetric tensor. All this leads to the notions of gauge fields and of gauge freedom
\be
\l{gf}
\cases{A\Rightarrow A+\partial\Lambda,&\cr
	F\Rightarrow F.&\cr}
\ee
So we can impose the gauge condition
\be
\l{L}
\partial.A=0,
\ee
which is, actually, an integrability condition for (\ref{iMa}) as it is then reduced to four uncoupled equations
\be
\l{we}
\Box A=-4\pi J.
\ee
The wave equation is indeed a basic and universal field equation in physics. This well known standard approach is common to all gauge fields but there are some questions that we had never thought of asking. It is clear, for example, that the gauge freedom is a consequence of the anti-symmetry of the interaction field but what is the reason that makes anti-symmetric all fundamental interaction fields? In General Relativity the potential, but not the field is described by a symmetric tensor. In this work we want to present a new approach that starting from the solutions of the wave equation (\ref{we}) obtain all the above equations (\ref{hM},\ref{iM},\ref{hs},\ref{iMa},\ref{gf},\ref{L}) as mere consequences that require no further assumption. The background in this new approach is endowed, of course, with more physical information than in the above standard one. It is based on the discrete field formalism, which is an alternative to the standard continuous one as a way of avoiding the basic cause of its ubiquitous problems with infinities: continuous fields, i.e. fields with infinite degrees of freedom.

Let us consider an spherical electromagnetic signal emitted from a point P, for fixing the idea. It propagates on the P-light-cone as a wave, distributed over an ever expanding sphere centred at P. It is a travelling spherical wave, an example of a continuous  field $A(x)$ defined with support on the lightcone. The field intensity decreases as the distance from the origin increases and, consequently, we have problems with an infinite energy-momentum density if we trace back this wave to P, its starting point where, let's say, we find a point electron. Infinite self-field is usually attributed to the assumption of a point-like source but recently it has been proved that this is not correct \cite{hep-th/9610028}. We know from quantum physics that this  field continuity is just apparent as this wave front is a collection of a, huge but finite, number of  field quanta. Now, in a new discrete approach to field theory, we want to see this classical spherical wave front as a collection of points, classical point-fields, to be defined below in a covariant way. Instead of studying the continuous field we study the discrete point-fields, and we find out that they are everywhere finite, have no singularity and propagate without changing themselves. Their connections to classical electrodynamics and to quantum physics are being discussed elsewhere \cite{ecwpd} in a more complete way. The standard continuous field (and all its problems) can always be retrieved by an integration over the point-fields; but we want to keep working with the point-like fields as the physically fundamental ones, considering the continuous one as just an approximation, a classical average field. We recur to causality for defining the discrete field in a consistent and manifestly covariant way.
\begin{center}
Causality: local and extended
\end{center}
\noindent Any given pair of events on Minkowski spacetime defines a four-vector $\Delta x.$ If this  $\Delta x$ is connected to the propagation of a free physical object (a signal, a particle, a field, etc) it is constrained to satisfy
\be
\label{1}
\Delta\tau^2=-\Delta x^{2}, 
\ee 
where $\tau$ is a real-valued parameter. We use a metric $\eta=diag(1,1,1,-1)$. So, (\ref{1}) just expresses that $\Delta x$ cannot be spacelike. A physical object does not propagate over a spacelike $\Delta x.$ This is {\it local causality}.  Geometrically it is the definition of a three-dimensional double cone; $\Delta x$ is the four-vector separation between a generic event $x^{\mu}\equiv({\vec x},t)$ and the cone vertex. See the Figure 1. This conic hypersurface, in field theory, is the free-field support: a {\bf free} field cannot be inside nor outside but only on the cone. The cone-aperture angle $\theta$ is given by $\tan\theta=\frac{|\Delta {\vec x}|}{|\Delta t|},\; c=1,$ or equivalently, $\Delta\tau^{2}=(\Delta t)^{2}(1-\tan^{2}\theta).$ A change of the supporting cone corresponds to a change of speed of propagation and is an indication of interaction.
Special Relativity restricts $\theta$ to the range $0\le\theta\le\frac{\pi}{4},$ which corresponds to a restriction on $\Delta\tau:$ $0\le|\Delta\tau|\le|\Delta t|.$ The lightcone ($\theta=\frac{\pi}{4},$ or $|\Delta\tau|=0$) and the t-axis in the object rest-frame ($\theta=0,$ or $|\Delta\tau|=|\Delta t|$) are the extremal cases.

For defining a discrete field a more restrictive constraint is required:
\be
\l{f}
\Delta\tau+  f.\Delta x=0,
\ee
where $f$ is defined by $f^{\mu}=\frac{dx^{\mu}}{d\tau},$ a constant  four-vector tangent to the cone; it is  timelike $(f^{2}=-1$) if $\Delta\tau\ne0,$  or lightlike $(f^{2}=0$) if $\Delta\tau=0$. \\
The equation (\ref{f}) defines a hyperplane tangent to the cone (\ref{1}). Together, (\ref{1}) and (\ref{f}) define a cone generator $f$, tangent to $f^{\mu}$. A fixed four-vector $f^{\mu}$ at a point represents a fiber in the spacetime, a straight line tangent to $f^{\mu}$, the $f$-generator of the local cone (\ref{1}).\\
Extended causality is the imposition of both (\ref{1}) and (\ref{f}) to the propagation of a signal, or of any physical object. Geometrically, it is a requirement that the signal remains on the cone generator $f$. 
\begin{center}
Discrete fields
\end{center}
\noindent We turn now to the question of how to define a field with support on a generic fibre $f$, a $(1+1)$-manifold embedded on a $(3+1)$-Minkowski spacetime.
As a consequence of the causality constraint (\ref{1}), the fields must be explicit functions of x and of $\tau,$ where $\tau$, is a supposedly known function of x, a solution of (\ref{1}):  $\tau=\tau_{0}\pm\sqrt{-(\Delta x)^{2}}.$ We have from (\ref{f}) that 
\be
\l{fmu}
f_{\mu}=-\frac{\partial\tau}{\partial x^{\mu}}.
\ee
For a massless field, as it propagates without a change on its proper time, $\Delta\tau=0$, $\;\tau$ is actually the instantaneous proper-time of its source at the event of its emission.  Well-known examples of this are the Lienard-Wiechert solutions, discussed in this context in \cite{hep-th/9610028}.\\
Let $A_{f}$ be the intersection of the wave front A  with the fibre $f.$ In other words,
\be
\label{Af}
A(x,\tau)_{f}=A(x,\tau){\Big |}_{f}.
\ee
The symbol $A_{f}$ or $A(x,\tau)_{f}$ denotes the restriction of $A(x,\tau)$ to the fibre f. It represents an element of $A(x,\tau)$, the part of it contained in the fibre $f$: a point propagating along $f.$ This definition (\ref{Af}) would not make any sense if the point character (discrete and localized) of $A_{f}$ could not be sustained during its time evolution governed by its wave equation. It is remarkable, as we will see, that it remains as a point-like field as it propagates.  On the other hand, $A(x,\tau)$ represents the collection of all such elements $A(x,\tau)_{f}$ from all possible fibres $f,$ and so the converse of (\ref{Af}) is given by
\be
\label{s1s}
A(x,\tau)=\frac{1}{2\pi}\int d^{4}f\;\delta(f^{2})A(x,\tau)_{f},
\ee
where 
\be
d^{4}f=df_{4}\;|{\vec f}|^{2}\;d|{\vec f}|\;d^{2}\Omega_{f}
\ee
and 
\be
\delta(f^{2})=\frac{1}{2|{\vec f}|}\{\delta(f^{4}-|{\vec f}|)+\delta(f^{4}+|{\vec f}|)\}.
\ee
We will see later that for the emitted $(f^{4}=|{\vec f}|)$ field in the source instantaneous rest frame at the emission time $(f^{4}=1)$ the equation (\ref{s1s}) can be written as
\be
\label{s}
A(x,\tau)=\frac{1}{4\pi}\int d^{2}\Omega_{f}A(x,\tau)_{f},
\ee
where the integral represents the sum over all directions of ${\vec f}$ on  a lightcone. $4\pi$, we see, is a normalization factor and (\ref{s}) is a particular case of
\be
\label{sss}
A(x,\tau)=\frac{\int_{\Omega_{f}} d^{2}\Omega_{f}A(x,\tau)_{f}}{\int_{\Omega_{f}} d^{2}\Omega_{f}}.
\ee
$A(x,\tau)$ is a continuous field and so $A(x,\tau)_{f}$ could also be seen as a continuous function of its parameter $f,$ but we prefer to draw another more interesting picture. We know from modern physics that this continuity of A is just an approximation. An electromagnetic wave, for example, is made of a large number of photons and, in a classical description, it could be seen as a bunch of massless point particles swarming out from their source in all directions. Although $A(x,\tau)_{f}$ is a field, a point-like field, we want to associate it to this classical idea of a point particle emitted, by its source, in the direction $f$ and we will call it the classical quantum (as contradictory as it may appear to be)  at the fibre $f.$ If A denotes an electromagnetic wave, $A_{f}$ will denote a classical photon. 

This represents a drastic change in the meaning of $A(x,\tau)$ and a reversal of what is taken as the primitive and the derived concept.  $A(x,\tau)$, while taken as the primitive concept, represented the actual physical field; now it has been  reduced to the average effect of a large, but finite, number of (classical) photons. $A(x,\tau)_{f}$ is a classical representation of the actual physical agent of electromagnetic interactions, the photons. $A(x,\tau)$ being, in this new context, just an average field representation of the exchanged photon can  produce good physical descriptions only at the measure of a large number of photons. This is the new interpretation of equation (\ref{s}). It does not make much difference for most of the practical situations that correspond, for example, to the description of an electromagnetic field involving a large number of photons and in a point far away from its sources. It certainly fails for very low intensity light involving few photons. Imagine an extreme case of just one emitted photon, pictured as the fibre $f$ in the Figure \ref{f2}.
Then $A(x,\tau),$  represented by the dotted circle, gives a false picture of isotropy while the true physical action (the photon) goes only along the $f$ direction. This situation is unavoidable when we approach to the field source, the electric charge, to such a small distance that the time elapsed between 
a photon emission and its detection is of the order of the period of time between two consecutive  (as they must be discrete) emissions of photons by the accelerated charge. In this extreme situation the description in terms of the averaged field $A(x,\tau)$, i.e. the usual field of electrodynamics , fails. The Gauss's equation has only meaning for a time-and-space-averaged electric field. This is the origin  of problems (infinite self-energy, the Lorentz-Dirac as an equation of motion, etc) for the usual local-causality-field theory  \cite{Rorhlich,Jackson,Teitelboim,Rowe,Lozada}, discussed, in this context, in \cite{hep-th/9610028} and \cite{ecwpd}. 
\begin{center}
Causality and dynamics
\end{center}
The constraint (\ref{fmu}) has a very important dynamical content as we discuss now. For a massless field emitted by a point charge on a worldline $z(\tau)$, parameterized by its proper-time $\tau$, $\Delta\tau=0$ and $\Delta x=x-z(\tau).$ The restriction (\ref{f}) is then reduced to $f.(x-z(\tau))=0$ and this implies that the event $x$, where the field is being observed, and the charge retarded position  $z(\tau)$ must belong to a same line $f$. It is not necessary to explicitly distinguish a generic $\tau$ from a $\tau$ at a retarded position, as the situations considered in this note always refers to the last one.
More information can be extracted from this constraint as $\partial_{\mu}f.(x-z){\Big |}_{f}=0$ implies, with (\ref{fmu}), on 
\be
\l{fv} 
f.V{\Big |}_{f}=-1,
\ee
where $V=\frac{dz}{d\tau}$.This relation  may be seen as a covariant normalization of the time component of $f$ to 1 in the electron rest-frame at its retarded time, 
\be
\l{f4}
f^{4}{\Big |}_{{f\;\;}\atop{{{\vec V}=0}}}=|{\vec f}|{\Big |}_{{f\;\;}\atop{{\vec V}=0}}=1.
\ee
This justifies the equation (\ref{s}). 
With a further derivation and with $a^{\mu}=\frac{dV^{\mu}}{d\tau}$ we get from (\ref{fv}) the following most important relationship
\be
\l{dA0}
a.f{\Big |}_{f}=0,
\ee
between the direction $f$ along which the signal is emitted (absorbed) and the instantaneous change in the charge state of motion at the retarded (advanced) time. It implies that
\be
\l{a4}
a_{4}=\frac{{\vec a}.{\vec f}}{f_{4}}{\Big|}_{f},
\ee
whereas $a.V\equiv0$ leads to
$a_{4}=\frac{{\vec a}.{\vec V}}{V_{4}}{\Big|}_{f},$
and so we have that in the charge instantaneous rest frame at the emission time ${\vec a}$ and ${\vec f}$ are orthogonal vectors,
\be
\l{af}
{\vec a}.{\vec f}{\Big |}_{{f\;\;}\atop{{\vec V}=0}}=0.
\ee
\begin{center}
Field equations
\end{center}
 \noindent The derivatives of $A_{f}(x,\tau),$ allowed by the constraint (\ref{f}), are the directional derivatives along $f,$ which with the use of (\ref{fmu}) we write as
\be
\label{fd}
\partial_{\mu}A_{f}=(\frac{\partial }{\partial x^{\mu}}+\frac{\partial \tau}{\partial x^{\mu}}\frac{\partial}{\partial \tau})A(x,\tau){\Big |} _{f}={\Big(}\frac{\partial }{\partial x^{\mu}}-f_{\mu}\frac{\partial}{\partial \tau}{\Big)}A_{f}\equiv\nabla_{\mu} A_{f}.
\ee
(\ref{1}) implies that $\tau$ is a function of $x$ whereas (\ref{f}) implies on (\ref{fmu}).
With $\nabla$ replacing $\partial$ for taking care of the constraint (\ref{f}), the propertime $\tau$ can be treated as a fifth independent  coordinate. We adopt this geometrical approach. The field equation for a massless field is
\be
\label{wef}
\eta^{\mu\nu}\nabla_{\mu}\nabla_{\nu}A_{f}(x,\tau)=J(x,\tau),
\ee
or, explicitly
$(\eta^{\mu\nu}\partial_{\mu}\partial_{\nu}-2f^{\mu}\partial_{\mu})A_{f}(x,\tau)=J(x,\tau),$
as $f^{2}=0$. $J$ is its source four-vector current.\\ 
An integration over the $f$ degrees of freedom in (\ref{wef}) reproduces, with the use of (\ref{s}), the usual wave equation of the standard formalism, 
 as $\int d^{2}\Omega_{f}f^{\mu}\partial_{\mu}\partial_{\tau}A_{f(x)}=0$ because $A_{f}(x)$ is an even function of $f$, as we can see from (\ref{pr9}) below. Observe that the missing $4\pi$ in (\ref{wef}) re-appears in (\ref{s}). So, the standard continuous formalism is retrieved from this discrete $f$-formalism with $A(x)$ as the average of $A_{f}(x)$, and (\ref{we}) as the average of (\ref{wef}), in the sense of (\ref{s1s}). 

The $f$-wave equation (\ref{wef}) can be solved by an f-Green's function,
\be
\label{sgf}
A_{f}(x,\tau_{x})=\int d^{4}yd\tau_{y}\; G_{f}(x-y,\tau_{x}-\tau_{y})\;J(y),
\ee
where the sub-indices specify the respective events $x$ and $y$, and $G_{f}(x-y,\tau_{x}-\tau_{y})$ being a solution of
\be
\label{gfe}
\eta^{\mu\nu}\nabla_{\mu}\nabla_{\nu}G_{f}(x-y,\tau_{x}-\tau_{y})=\delta^{4}(x-y)\delta(\tau_{x}-\tau_{y}):=\delta^{5}(x-y).
\ee
This equation has been solved in \cite{ecwpd}:
\be
\label{pr9}
G_{f}(x,\tau)=\frac{1}{2}\theta(-b{\bar f}.x)\theta(b\tau)\delta(\tau+  f.x)=\frac{1}{2}\theta(bf^{4}t)\theta(b\tau)\delta(\tau+  f.x),
\ee
where $b =\pm1,$ and $\theta (x)$ is the Heaviside function, $\theta(x\ge0)=1$ and $\theta(x<0)=0.$  $G_{f}(x,\tau)$ does not depend on ${\vec x}_{\hbox {\tiny T}}$, where the subscript ${\hbox {\tiny T}}$ stands for transversity with respect to ${\vec f},$: ${\vec f}.{\vec x}_{{\hbox {\tiny T}}}=0.$  This justifies our approach of working with discrete field; its propagation does not depend on ${\vec x}_{\hbox {\tiny T}}$, or in other words, on the parts of $A(x)$ that are not in $f$. So, each $A_{f}$ can be treated as an independent single physical entity.\\
The properties of $G_{f}$ are discussed in \cite{ecwpd}. For $f^{\mu}=({\vec f}, f^{4})$, ${\bar f}$ is defined by ${\bar f}^{\mu}=(-{\vec f}, f^{4});$ $f$ and ${\bar f}$ are two opposing generators of a same lightcone; they are associated, respectively, to the $b=+1$ and to the $b=-1$ solutions and, therefore, to the processes of creation and annihilation of a photon.
For $b=+1$ or $t>0$, $G_{f}(x,\tau)$ describes a point signal emitted by the electron  at $\tau_{ret}=0,$ and that has propagated to $x$ along the fiber $f,$ of the future lightcone of $z(\tau_{ret})$;  for $b=-1$ or $t<0,$  $G_{f}(x,\tau)$ describes a point signal that is propagating  along the fiber $\bar{f}$ of the future lightcone of $x$ towards the point $z(\tau_{adv})$ where it will be absorbed (annihilated) by the electron. See the Figure \ref{f3}.
Observe the differences from the standard interpretation of the Li\`enard-Wiechert solutions. There is no advanced , causality violating solution here. $J$ is the source of the $f$ solution and a sink for the $\bar{f}$ solution. These two solutions correspond to creation and annihilation of particles, of classical photons. $G_{f}$ has no singularity, in contradistinction to the standard Green function
\be
\l{sg} 
G(x,\tau)=\frac{1}{r}\Theta(bt)\delta(r-bt),
\ee
 which is retrieved \cite{ecwpd} by with use of (\ref{s1s}). So the discrete field $A_{f}$ doesn't know any singularity in the sense of infinities. The singularity presented by its average $A(x)$ is a consequence of this averaging process on its definition (\ref{s1s}). This is also true for the gravitational field of General Relativity, as discussed in \cite{gr-qc/9801040}. 
\begin{center}
Gauge fields
\end{center}
Let us now apply this $f-$formalism to the electromagnetic field generated by a classical point spinless electron, which does not imply on losing generality. The essence of the discrete formalism is that fields and sources are discrete, point-like. The continuous image is just a macroscopic approximation, an average. In this formalism where $\tau$ is treated as an independent fifth parameter, a definition of a four-vector current must carry an additional constraint expressing the causal relationship between two events $y$ and $z$. Its four-vector current is given by
\be
\l{J}
J^{\mu}(y,\tau_{y}=\tau_{z})= eV^{\mu}(\tau_{z})\delta^{3}({\vec y}-{\vec z})\delta(\tau_{y}-\tau_{z}),
\ee
where $z^{\mu}(\tau_{z}),$ is the electron worldline parameterized by its proper-time $\tau_{z}.$ In (\ref{J}) $\tau_{y}$ has to be equal to $\tau_{z}$ as a consequence of the Dirac deltas and of (\ref{1}). For $b=+1$, that is, for the field emitted by J we have
\be
A_{f}(x,\tau_{x})=2e\int d^{5}y G_{f}(x-y)V^{\mu}(\tau_{y})\delta^{3}({\vec x}-{\vec y})\delta(\tau_{x}-\tau_{y}),
\ee
where the factor 2 accounts for a change of normalization with respect to (\ref{sgf}) as we are now excluding the annihilated photon (the integration over the future lightcone). Then,
\be
\l{Af1}
A_{f}(x,\tau_{x})=eV^{\mu}(\tau_{z})\theta(t_{x}-t_{z})\theta(\tau_{x}-\tau_{z}){\Big |}_{{\tau_{z}=\tau_{x}+f.(x-z)}}.
\ee
So, the field $A_{f}$ is given, essentially, by the charge times its four-velocity at its retarded time. $\nabla\theta(t)$ and $\nabla\theta(\tau)$ do not contribute to $\nabla A_{f},$ except at $x=z(\tau),$ as a further consequence of the field constraints. So, for $\tau=0$ and $t>0$ we write just 
\be
\l{dAf}
\nabla_{\nu}A^{\mu}_{f}=\nabla_{\nu}(eV^{\mu}){\Big |}_{f}=-ef_{\nu}a^{\mu}{\Big |}_{f}.
\ee
The Lorentz gauge condition $\nabla.A_{f}=0,$ is then a consequence of (\ref{dA0}) and so it is automatically satisfied by the field $A_{f}$ of (\ref{Af1}). The extended causality constraint, i.e. the explicit constraint of a causal propagation of the field, whichever be the field, leads to the constraint (\ref{dA0}) between the field direction of propagation $f$ and the change in the state of motion of its source (sink) at the emission (absorption) time. $A_{f}$, given by (\ref{Af1}), solution to (\ref{wef}) for a point charge $e$, is just an expression of the ``charge state of motion" at the emission time. But, ``state of motion" or velocity is a relative or frame dependent concept and cannot, by itself, be used to define or characterize a physical object; the acceleration, an absolute concept, can do that. Then $A_{f}$ must be seen as a potential, and  the physical field must be associated to its gradient. For writing a Lagrangian, we can use $A_{f}$ for the potential field as we do with the velocity or momentum in classical mechanics, but only to the field is attributed the physical meaning of force carrier. Thus, we see from (\ref{dAf}) and (\ref{af}) that the physical field is given by a second-rank tensor 
\be
F^{\mu\nu}_{f}\Leftrightarrow-\nabla^{\nu}A_{f}^{\mu}=ef^{\nu}a^{\mu}{\Big |}_{f},
\ee
defined by the acceleration of its source at its emission time, but in such a way that in the charge instantaneous rest-frame (see (\ref{af}) this dependence is reduced to 
\be
\l{ft}
F_{f}\Leftrightarrow efa_{\hbox{\tiny T}}{\Big |}_{f}=ef(a-f\frac{{\vec a}.{\vec f}}{f_{4}^2}){\Big |}_{f},
\ee
 as $a_{\hbox{\tiny T}}:=a-f\frac{{\vec a}.{\vec f}}{f_{4}^2}.$
 The field $F_{f}$ could, in principle, be a symmetric or an anti-symmetric tensor but we see from (\ref{ft}) that Lorentz covariance requires anti-symmetry. So, 
\be
\l{FedA}
F^{f}_{\mu\nu}:=-e(f_{\mu}a_{\nu}-f_{\nu}a_{\mu}){\Big |}_{f}=\nabla_{\mu}A^{f}_{\nu}-\nabla_{\nu}A^{f}_{\mu},
\ee
which leads us to the Maxwell's theory on a fibre $f$. Therefore we conclude that causality and Lorentz covariance impose an anti-symmetry and, therefore, a gauge freedom, on the interaction fields. This is also true for General Relativity \cite{gr-qc/9801040}. So we can understand why there is no symmetric rank two tensor field describing fundamental interactions. The field singularities are consequences of their definition as spacetime average fields. Details and extended discussions on the physical interpretation of $F_{f}$, on its connections to classical and quantum physics and to experimental data are being presented in \cite{ecwpd}.


\vglue-1cm 

\vglue2cm 
\hspace{-3cm}
\vglue3cm

\hspace{-4.5cm}
\parbox[t]{5.0cm}{
\begin{figure}
\vglue-3cm
\epsfxsize=400pt
\epsfbox{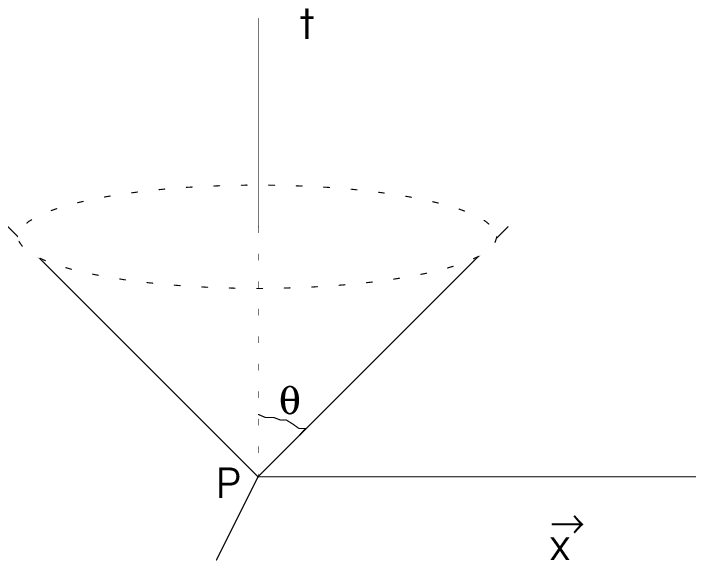}
\vglue-12cm
\caption{}
\l{f1}
\end{figure}}\hfill
\\ \mbox{}
\hspace{5cm}
\parbox[t]{8.0cm}
{\vglue-6cm 
 Local causality.\\ The relation $\Delta\tau^2=-\Delta x^{2},$ a causality constraint, is seen as a restriction of access to regions of spacetime. It defines a three-dimension cone which is the spacetime available to a point-like, free, physical object at the cone vertex. The object is constrained to be on the cone.}\\ \mbox{}
\vglue-1cm

\vglue-10cm

\parbox[]{7.5cm}{
\begin{figure}
\vglue6cm
\epsfxsize=400pt
\epsfbox{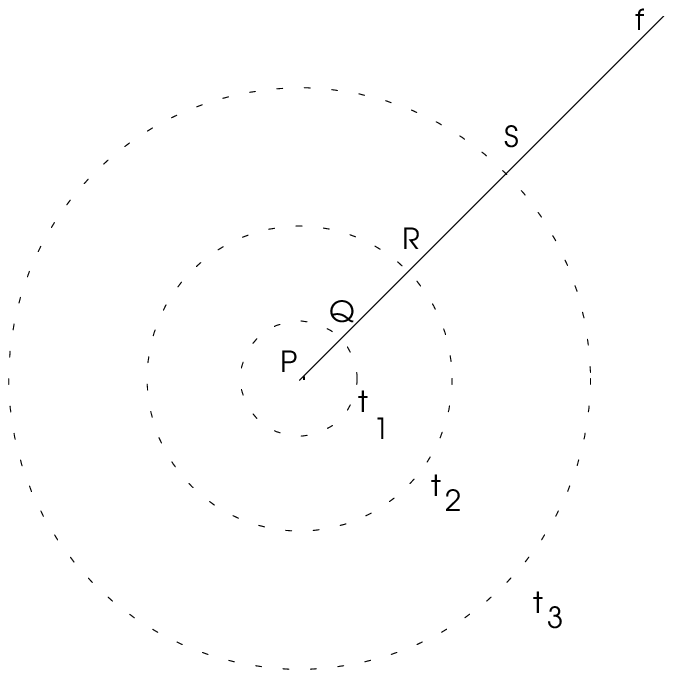}
\vglue-6cm
\caption{}
\l{f2}
\end{figure}
\vglue-0cm}\\
\mbox{}
\hfill
\hspace{5.0cm}
\parbox[]{7.5cm}{\vglue-8cm
The relationship between the fields $A_{f}$ and $A$. The three doted circles represent, at three instants of time, the field $A$ as an spherically symmetric signal emitted by a charge at the point P. The straight line PQRS\dots is the fibre $f,$ a lightcone generator tangent to $f^{\mu}.$ The points Q, R, and S are a classical photon $A_{f}$ at three instants of time. In the case of a very low intensity light with just one photon the field $A$ transmit a false idea of isotropy. }\\ \mbox{}

\vglue-5cm

\vglue2cm 
\hspace{-3cm}
\vglue3cm

\hspace{-1cm}
\parbox[t]{5.0cm}{
\begin{figure}
\vglue-3cm
\epsfxsize=400pt
\epsfbox{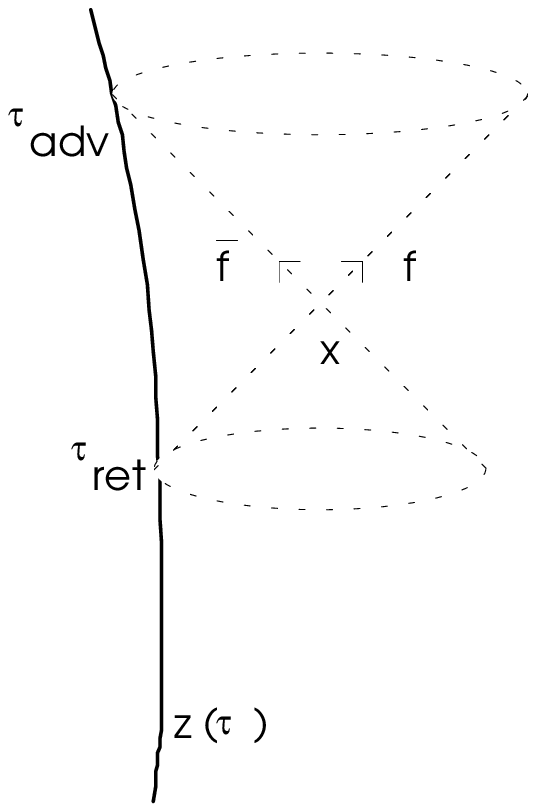}
\vglue-6.5cm
\caption{}
\l{f3}
\end{figure}}\hfill
\\ \mbox{}
\hspace{5cm}
\parbox[t]{8.0cm}
{\vglue-6cm Creation and annihilation of particles.\\
 The Li\`enard-Wiechert solutions as creation and annihilation of particles. There are two classical photons at the point x: one, created at $\tau_{ret}$, has propagated to x on the lightcone generator f; the other one propagating on the lightcone generator ${\bar f}$ from x towards the electron worldline where it will be annihilated at $\tau_{adv}$.}\\ \mbox{}
\vglue-1cm

\end{document}